\documentclass{aa}  
\usepackage{graphicx}
\usepackage{txfonts}
\newcommand{\oc}{$O\!-\!C$}
\newcommand{\ova}{Ostrava}
\newcommand{\valmez}{Vala\v{s}sk\'e Mezi\v{r}\'{\i}\v{c}\'{\i}}

\begin{document}
\title{DI~Her as a test of internal stellar structure and General Relativity}

\subtitle{New apsidal motion rate and evolutionary models}

\author{A. Claret \inst{1}
\and G. Torres \inst{2}
\and M. Wolf \inst{3}
}

\offprints{A. Claret, e-mail:claret@iaa.es}

\institute{Instituto de Astrof\'\i sica de Andaluc\'\i a, CSIC,
           Apartado 3004, E-18080 Granada, Spain \and
           Harvard-Smithsonian Center for Astrophysics, 60 Garden St.,
           Cambridge, MA 02138, USA \and Astronomical Institute,
           Faculty of Mathematics and Physics, Charles University
           Prague, CZ-180~00 Praha 8, V~Hole\v{s}ovi\v{c}k\'ach 2,
           Czech Republic}

\date{Received; accepted 2010 February 9}

\abstract
{For the past three decades, and until recently, there has been a
serious discrepancy between the observed and theoretical values of the
apsidal motion rate $\dot\omega$ of the eccentric eclipsing binary
DI~Her, which has even been interpreted occasionally as a possible
failure of General Relativity (GR). A number of plausible explanations
have been put forward.  Recent observations of the Rossiter-McLaughlin
effect have shown convincingly that the reason for the anomaly is that
the rotational axes of the stars and the orbital axis are misaligned,
which changes the predicted rate of precession significantly.}
{Although as a result of those measurements the disagreement is now
drastically smaller, it remains formally at the level of 50\%,
possibly due to errors in the measured apsidal motion rate, outdated
stellar models, or inaccuracies in the stellar parameters. The aim of
this paper is to address each of these issues in order to improve the
agreement further.}
{New times of minimum have been collected, and used for a
redetermination of the apsidal motion rate. Based on the latest
determinations of the absolute dimensions of the binary, we have
computed new stellar evolution models with updated physical inputs,
and derived improved apsidal motion constants for the components. We
have performed Monte Carlo simulations to infer the theoretical
distribution of $\dot\omega$, including the contributions from GR as
well as tidal and rotational distortions. All observational errors
have been accounted for.}
{Our simulations yield a retrograde apsidal motion rate due to the
rotationally-induced oblateness of $-0.00056$ deg~cycle$^{-1}$ (mode
of the distribution), a GR contribution of $+0.00068$
deg~cycle$^{-1}$, and a tidal contribution of $+0.00034$
deg~cycle$^{-1}$, leading to a total predicted rate of $+0.00046$
deg~cycle$^{-1}$.  This is in excellent agreement with the newly
measured value of $+0.00042$ deg~cycle$^{-1}$. The formal difference
is now reduced to 10\%, a small fraction of the observational
uncertainties.}
   {}

\maketitle

%

\section{Introduction}
\label{sec:introduction}

For some three decades the eccentric B-type eclipsing binary DI~Her
($P = 10.55$ days) has been known to have an apsidal motion rate that
is too slow compared to theoretical predictions (for a brief review,
see Claret 1998). Until very recently this presented a serious
challenge to our understanding of stellar physics, and even the
validity of General Relativity has been called into question given
that the GR contribution is dominant for DI~Her. The discrepancy in
the total rate of precession amounts to 400\% or more. This anomaly
persisted even when considering hypothetical stellar models with
infinite mass concentration (apsidal motion constant $k_2 = 0$), which
tends to reduce the predictions, and this was often interpreted as a
failure of General Relativity. The problem was all the more puzzling
when considering that studies of other binary systems suitable for
testing GR and with well determined absolute dimensions indicated good
agreement with theory (see, e.g., Claret 1997 and Wolf et al.\
2009). This suggested to many that the problem may lie with DI~Her
itself. A number of hypotheses were put forward over the years in
order to explain the disagreement, including
\begin {enumerate}
\item A rapid ongoing circularization of the orbit, affecting the
   measured times of minimum;
\item The presence of a circumstellar cloud between the components,
   which would affect the gravitational field of the stars;
\item The presence of a distant third star in the system, which would
   contribute to the motion of the line of apsides;
\item An alternative theory of gravitation.
\end {enumerate}
Some of these ideas were examined quantitatively and ruled out by
Claret (1998). The third body hypothesis has been a serious contender
for many years (see, e.g., Khodykin 2007), although no spectroscopic
or photometric evidence of it has ever been found.  Difficulties in
the measurement of such a small apsidal motion as that of DI~Her
($\dot\omega_{\rm obs} = 1.04^{\circ}$ per 100 yr, or 0.00030
deg~cycle$^{-1}$; Guinan et al.\ 1994) were also suggested as a
possible explanation from time to time. 

An attractive alternative is a configuration for the binary in which
the spin axes are tilted relative to the orbital axis (Shakura 1985,
Guinan \& Maloney 1985, Company et al.\ 1988, Claret 1998). This would
introduce a correction to the rotational term of the apsidal motion
that can be negative (regression of the line of apsides), resulting in
a smaller total $\dot\omega$. The young age of the system (a few Myr)
certainly makes this plausible, as tidal forces would not have had
time to operate and bring the spin axes of the stars into alignment
with the axis of the orbit (see Sect.~\ref{sec:models}).  A remarkable
confirmation of this prediction was obtained recently by Albrecht et
al.\ (2009), who provided compelling observational evidence for a
peculiar orientation of the stars by taking advantage of the
Rossiter-McLaughlin effect. This is a distortion of the spectral lines
that occurs during the eclipses, as a result of the partial
obscuration of the rotating stellar disks.  Using this effect, they
were able to measure the angles between the sky projections of the
spin axes and the orbital axis, and show that there is indeed a strong
misalignment for both components of the binary. It is the first
documented case in which this has been shown convincingly for stars in
a double-lined eclipsing binary.\footnote{Spin-orbit misalignment has
been reported previously for several host stars of extrasolar
transiting planets.}

With this new information Albrecht et al.\ (2009) managed to reduce
the discrepancy with the predicted apsidal motion rate of DI~Her very
significantly, so that theory and observation are now in much better
agreement.  Some of the angles involved in computing the effect of the
tilted axes remain unknown, however, so the theoretical prediction
necessarily assumes a statistical character in this case, which
Albrecht et al.\ (2009) tackled by means of Monte Carlo simulations.
The nominal difference between the measured apsidal motion rate and
the mode of their theoretical probability distribution for
$\dot\omega$ is still about 50\% at present, although the distribution
is rather wide so theory and observation are formally consistent with
each other. Some of this difference could be due to the measurement of
$\dot\omega_{\rm obs}$ itself, to inaccuracies in the adopted absolute
dimensions of the stars, or to the stellar evolution models used to
infer the apsidal motion constants $k_2$, which are needed as inputs
to the theoretical calculations.

In this paper we re-examine all the ingredients in order to provide
the most accurate predictions and observational constraints. We
incorporate new measurements of the times of eclipse to derive an
improved value of $\dot\omega_{\rm obs}$ for DI~Her that is
significantly larger than the value adopted by Albrecht et al.\
(2009). We use the most recent determinations of the mass, radius, and
other properties of the stars, and we revisit the temperature
determinations. With these we provide new fits to stellar evolution
models and derive new apsidal motion constants. We then repeat the
comparison of the measured apsidal motion rate with theory, also
examining the sensitivity of the predictions to several of the key
observables.

\section{Observational data}
\subsection{The absolute dimensions}
\label{sec:dimensions}

The absolute masses and radii of the components of DI~Her used here
are adopted from the recent compilation by Torres et al.\ (2009a), in
which these and other properties were re-examined or redetermined with
a uniform methodology. The uncertainties are adopted from the same
source. We summarize all observational results in
Table~\ref{tab:dimensions}, and describe some of them later.  The
effective temperatures reported by Torres et al.\ (2009a) are based on
the individual $B\!-\!V$ colors as given by Popper (1982), and on his
color/temperature relation (Popper 1980). As a check, we used the
Str\"omgren photometry obtained by Hilditch \& Hill (1975) and the
calibration of Crawford (1978), and derived a mean de-reddened color
for the binary that is essentially identical to Popper's. The
color/temperature calibration by Paunzen et al.\ (2005), which
combines several more recent temperature scales, then yields
individual temperatures of 17300~K and 15400~K for the primary and
secondary, with uncertainties of 800~K.  These are the temperatures we
adopt in this work, which are 300~K hotter than those listed by Torres
et al.\ (2009a).

\begin{table}
\centering          
\caption{Observational parameters for DI~Her.}
\label{tab:dimensions}      
\begin{tabular}{l c c}     
\hline\hline\noalign{\smallskip}
Parameter                                 &  Primary            &  Secondary         \\
\noalign{\smallskip}\hline\noalign{\smallskip}
Mass $m$ (M$_{\odot}$)                    &  $5.17 \pm 0.11$   &  $4.524 \pm 0.066$ \\
Radius $R$ (R$_{\odot}$)                  &  $2.681 \pm 0.046$  &  $2.478 \pm 0.046$ \\
Effective temperature (K)                 &  $17300 \pm 800$\phantom{22} &  $15400 \pm 800$\phantom{22}   \\
$v \sin\beta$ (km~s$^{-1}$)               &  $108 \pm 4$\phantom{22}    &  $116 \pm 4$\phantom{22}  \\
$\lambda$ (deg)                           &  $+72 \pm 4$\phantom{$+2$}   &  $-84 \pm 8$\phantom{$-2$}       \\
$\log k_2$                                &  $-2.10 \pm 0.05$\phantom{$-$}   &  $-2.13 \pm 0.05$\phantom{$-$}  \\
\noalign{\smallskip}\hline\noalign{\smallskip}
Orbital period $P$ (days)*                & \multicolumn{2}{c}{$10.5501696 \pm 0.0000007$\phantom{2}} \\
Reference epoch $T_0$ (HJD)               & \multicolumn{2}{c}{$2,\!447,\!372.9567 \pm 0.0005$\phantom{$2,\!447,\!37$}} \\
$\omega_0$ at reference epoch (deg)       & \multicolumn{2}{c}{$330.0 \pm 0.1$\phantom{33}} \\
Orbital eccentricity $e$                  & \multicolumn{2}{c}{$0.4895 \pm 0.0008$} \\
Orbital inclination $i$ (deg)             & \multicolumn{2}{c}{$89.30 \pm 0.07$\phantom{2}} \\
$\dot\omega_{\rm obs}$ (deg cycle$^{-1}$) & \multicolumn{2}{c}{$0.00042 \pm 0.00012$} \\
\noalign{\smallskip}\hline\noalign{\smallskip}
\end{tabular}

* $P$ corresponds here to the sidereal period $P_s$.
\end{table}

\subsection{New apsidal motion rate}
\label{sec:wolf}

The apsidal motion rate of DI~Her was determined by means of an \oc\
diagram analysis.  The method we followed is that described by
Gim\'enez \& Garc\'{\i}a-Pelayo (1983), which is an iterative weighted
least-squares procedure including terms in the eccentricity up to the
fifth order.  There are five adjustable quantities to be determined:
the rate of periastron advance $\dot{\omega}$, the sidereal period
$P_s$, the eccentricity $e$, and the longitude of periastron
$\omega_0$ at the reference epoch $T_0$.  The periastron position $\omega$
at any epoch $E$ is then given by the linear equation
\begin{equation}
\omega = \omega_0 + \dot{\omega}\ E~,
\end{equation}
and the relation between the sidereal and the anomalistic periods,
$P_s$ and $P_a$, is given by
\begin{equation}
P_s = P_a \,(1 - \dot{\omega}/360^\circ)~.
\end{equation}
The value for the orbital inclination was adopted as $i = 89\fdg30$
from the light-curve analysis of Guinan \& Maloney (1985).

\begin{figure}
\includegraphics[width=0.49\textwidth]{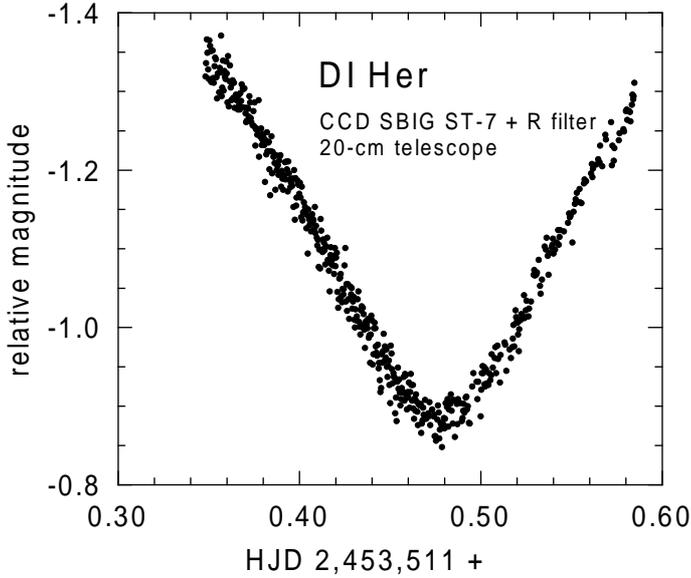}
\caption{Differential light curve of DI Her during the primary minimum
of 2005 May 20, as measured by Ms.\ Hana Ku\v{c}\'akov\'a at the
Johann Palisa Observatory, Technical University Ostrava, Czech
Republic.  The observations were made through a Johnson $R$ filter
with a 20-cm reflecting telescope and SBIG ST-8XME CCD camera, with
exposure times of 25 seconds.  The comparison star was GSC~02109-00247
($V = 10.04$).  The light curve contains more than 500 points obtained
during nearly 6 hours of monitoring.}
\label{fig:minimum}
\end{figure}

We used all precise photoelectric and CCD timings recently
recalculated by Kozyreva \& Bagaev (2009, their Tables~1 and 2), as
well as new times of eclipse obtained by us at two observatories, with
the following instrumentation:
\begin{description}
\item{$\bullet$} Observatory and Planetarium of Johann Palisa, \ova,
Czech Republic: 0.20-m or 0.30-m telescopes with an SBIG~ST-8XME CCD
camera and Johnson $V$, $R$, and $I$ filters;
\item{$\bullet$} \valmez\ Observatory, Czech Republic: 0.30-m Celestron
Ultima telescope with an SBIG~ST-7 CCD camera and Johnson $R$ filter.
\end{description}
The reduction of all images was performed with the synthetic aperture
photometry software developed by Motl (2007).  The frames were
dark-subtracted and flat-fielded, and the heliocentric correction was
applied. An example of these observations is presented in
Figure~\ref{fig:minimum}. The new times of primary and secondary
eclipse and their errors were determined by the classical Kwee \& van
Woerden (1956) algorithm, and are listed in Table~\ref{newmin}.
Earlier visual and photographic times were not used in our analysis
because of the large scatter of those measurements.  A total of 69
times of minimum light were employed, including 34 corresponding to
the secondary eclipse.

\begin{figure}
\includegraphics[width=0.49\textwidth]{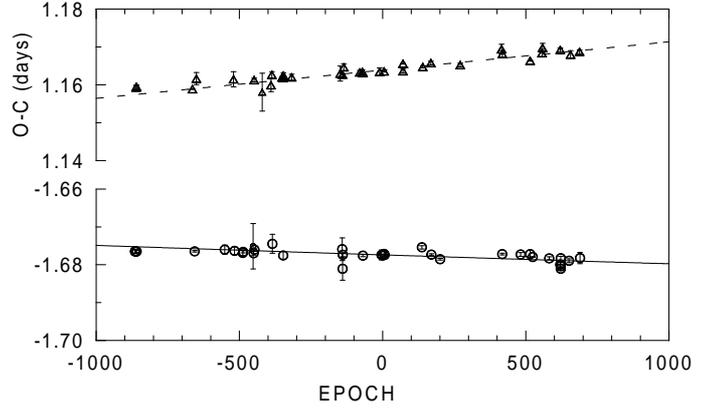}
\caption{The $O\!-\!C$ diagram of DI~Her (eclipse timing residuals),
together with our best-fit apsidal motion model.  The solid line and
circles correspond to the primary, and the dashed line and triangles
to the secondary.}
\label{fig:omc}
\end{figure}

\begin{figure}
\includegraphics[width=0.49\textwidth]{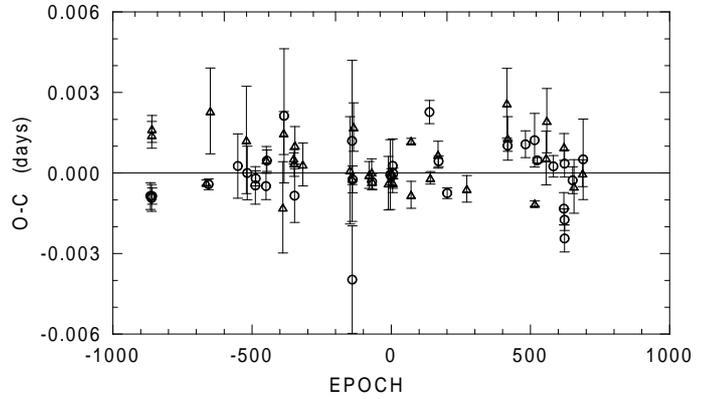}
\caption{The $O\!-\!C$ residuals from the photoelectric and CCD
timings of DI~Her, after subtraction of the apsidal motion
terms. Symbols are as in Figure~\ref{fig:omc}.}
\label{fig:residuals}
\end{figure}

The solution was performed by minimizing the standard $\chi^2$ sum,
and the errors of the observations were adopted as published. The
resulting apsidal motion parameters and their uncertainties from the
least squares fitting are presented at the bottom of
Table~\ref{tab:dimensions}, and the corresponding \oc\ diagram is
shown in Figure~\ref{fig:omc}.  The $\chi^2$ value for the best fit
was $\chi^2 \simeq 500$. This is considerably larger than the number
of degrees of freedom ($\nu = 64$), which is most likely due to the
formal errors of the eclipse timings being underestimated, as is often
found to be the case.  The uncertainty of $\dot\omega$ reported in
Table~\ref{tab:dimensions} already accounts for this effect.  The
extra scatter compared to the uncertainties is seen more clearly in
Figure~\ref{fig:residuals}, in which the influence of the apsidal
motion terms has been subtracted from the measurements. There is no
evidence in this diagram of any other significant variations, such as
might be induced by the presence of a third body in the system.

\begin{table}
\centering
\caption{New times of minimum light of DI~Her.}
\label{newmin}
\begin{tabular}{llrcl}
\hline\hline\noalign{\smallskip}
 HJD$-2,\!400,\!000$    &  $\sigma$ (days)  & Epoch  & Filter & Observatory  \\
\noalign{\smallskip}\hline\noalign{\smallskip}
52456.46107  & 0.0005 & 482.0 &  $R$  &  \ova \\
53511.47699  & 0.0004 & 582.0 &  $R$  &  \ova \\
53933.48379  & 0.0005 & 622.0 &  $I$  &  \valmez  \\
54239.43802  & 0.0008 & 651.0 &  $R$  &  \ova  \\
54627.36687* & 0.0005 & 687.5 & $VRI$ &  \ova  \\
\noalign{\smallskip}\hline\noalign{\smallskip}
\end{tabular}

* Mean value of $VRI$ measurements.
\end{table}

~\vskip 0.5truein

\section{Stellar evolution models}
\label{sec:models}

As a prerequisite for investigating the theoretical apsidal motion of
DI~Her, the observed stellar properties such as the mass, radius, and
temperature of each component must be well matched by stellar
evolution models, at a single age for the binary, since these models
are then used to infer the apsidal motion constants $k_2$ that factor
into the rotational and tidal contributions to $\dot\omega$. The basic
properties of the stellar evolution code used here have been described
by Claret (2004). The models include overshooting beyond the formal
boundary of the convective core, as determined by the Schwarzschild
criterion. The distance over which this is considered to occur is
specified as d$_{\rm over}$ = $\alpha_{\rm ov} $H$_p$, where H$_p$ is
the pressure scale height taken at the edge of the convective core, as
given by Schwarzschild's criterion, and $\alpha_{\rm ov}$ is a free
parameter. We adopt here $\alpha_{\rm ov} = 0.20$, following the
results of Claret (2007).  The evolution code incorporates radiative
opacities from the tables provided by Iglesias \& Rogers (1996), and
the calculations of Alexander \& Ferguson (1994) for lower
temperatures.

\begin{figure}
\includegraphics[width=0.48\textwidth]{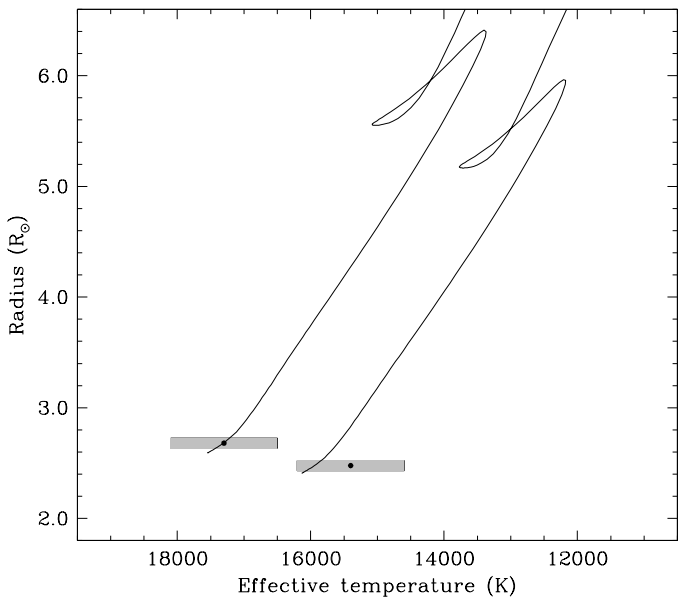}
\caption{Stellar evolution models compared with the measurements for
DI~Her. The evolutionary tracks are computed for the measured masses
(see Table~\ref{tab:dimensions}). The best fit has ($X$, $Z$) = (0.71,
0.02), implying a chemical composition near solar.}
\label{fig:tracks}
\end{figure}

\begin{figure}
\includegraphics[width=0.48\textwidth]{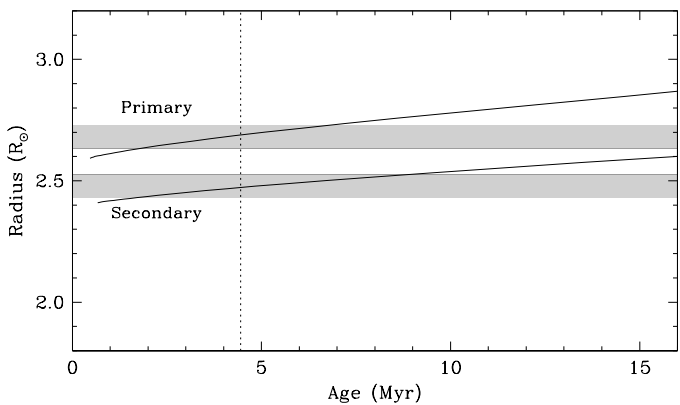}
\caption{Radius evolution of DI~Her, based on the models shown in
Figure~\ref{fig:tracks}. The shaded areas represent the measured radii
and their uncertainties.  The best-fit age of about 4.5~Myr is
indicated by the vertical dotted line.}
\label{fig:age}
\end{figure}

The chemical composition of double-lined eclipsing binaries is often
difficult to determine observationally, and that is the case also for
DI~Her.  This introduces some uncertainty in the comparison with
stellar evolution models, since the metal abundance $Z$ then becomes a
free parameter. For young systems such as DI~Her, $Z$ is quite
sensitive to the effective temperatures of the stars. We computed
evolutionary tracks for the measured masses and explored a range of
compositions, finding the best fit for ($X$, $Z$) = (0.71, 0.02),
which suggests an abundance very near that of the Sun. The
observations along with our best fit are displayed in
Figure~\ref{fig:tracks}. The agreement with the temperatures is quite
satisfactory, given the error bars and other uncertainties described
below.  Precise age determinations for young systems are not easy to
make and require very accurate measurements of the absolute
radii. This is illustrated clearly in Figure~\ref{fig:age}. As small
as the radius errors already are in DI~Her ($< 2$\% relative errors,
which are among the best available for eclipsing binaries; see Torres
et al.\ 2009a), the uncertainties allow for a wide range of ages
between $\sim$2~Myr and $\sim$7~Myr.  Nominally the best fit to the
radii is achieved at about 4.5~Myr with these models. We note, in
passing, that other models such as those from the Yonsei-Yale series
(Yi et al.\ 2001, Demarque et al.\ 2004) do not yield as good a fit,
at least with the grid of standard compositions ($X$, $Z$) that are
publicly available.

The theoretical apsidal motion constants inferred from our models are
$\log k_{21} = -2.10 \pm 0.05$ and $\log k_{22} = -2.13 \pm 0.05$ for
the primary and secondary, respectively. These values indicate that
the modern models favor stars that are slightly more centrally
concentrated in mass than those adopted by Claret (1998) and Albrecht
et al.\ (2009). This goes in the direction of making the predicted
apsidal motion rate smaller. Tests were made in order to evaluate the
influence on the inferred $k_2$ of uncertainties in the age
determination and in the chemical composition, and these uncertainties
are included in the errors reported above.

Regarding the measured effective temperatures, it is worth mentioning
that because the spin axes of the stars are tilted relative to the
orbital axis, the temperature distribution over their disks due to
gravity darkening could be highly non-uniform and asymmetric as viewed
from the Earth, and it is even possible one or both components are
viewed nearly pole-on. There may therefore be a bias in the measured
effective temperatures that could amount to several hundred degrees
for DI~Her, but because the exact orientation is not known (see next
section), the effect is difficult to estimate more precisely. Rotation
itself changes the internal structure of stars and affects their
evolution, generally leading to cooler temperatures. To explore this
effect more quantitatively for DI~Her we have computed models
accounting for rotation (rigid-body approximation), and we find
differences of the order of 200~K, depending on the inclination
relative to the line of sight. Once again, since the precise
orientation of the spin axes is unknown, we are unable to be more
specific regarding this effect.

With the models computed in this section we have calculated the age at
which the stellar spin axes in DI~Her are expected to become aligned
with the axis of the orbit due to the action of tidal forces. The
procedure follows closely that described by Torres et al.\ (2009b, and
references therein) for the case of the eclipsing system LV~Her. The
result using the theory by Zahn (1977, 1989) is $\sim$10$^8$~yr.  With
the more general equations by Hut (1981), valid for systems of high
eccentricity such as this, we obtain $3\times 10^7$~yr, which is an
order of magnitude larger than the evolutionary age. These estimates
suggest, as is observed, that tidal forces have not had sufficient
time in this system to make the rotation axes parallel to the axis of
the orbit.

\section{The apsidal motion of DI~Her}
\label{sec:apsidalmotion}

The total rate of apsidal motion from theory, to be compared with the
measured value, is given by sum of the individual contributions of
each component due to tidal and rotational distortions, and the
general relativistic term. When accounting for the possibility of
misalignment (Shakura 1985), we have
\begin{equation}
 \dot\omega_{\rm total} = \dot\omega_{\rm GR}+\dot\omega_{\rm tidal, 1}
 +\dot\omega_{\rm tidal, 2} + \dot\omega_{\rm rot, 1}\thinspace \phi_1+
 \dot\omega_{\rm rot, 2}\thinspace \phi_2~,
\label{eq:wtot}
\end{equation}
where
\begin{equation}
\dot\omega_{{\rm tidal,} j} = \left[{15 m_{3-j}\over{m_{j}}}k_{2j}\thinspace
g(e)\right]{\left({R_{j}\over{A}}\right)^{5}}~,
\label{eq:wtidal}
\end{equation}
\begin{equation}
\dot\omega_{{\rm rot,} j} = \left[
\left({\Omega_{j}\over{\Omega_K}}\right)^{2} \left(1 +
{m_{3-j}\over{m_{j}}}\right)f(e)\thinspace k_{2j}
\right]{\left({R_{j}\over{A}}\right)^{5}}~,
\label{eq:wrot}
\end{equation}
%
%
\begin{equation}
\phi_j = - {1\over{\sin^2 i}}\left[\cos \alpha_j\thinspace (\cos \alpha_j - \cos \beta_j \cos i)
+ {1\over{2}} \sin^2 i\thinspace (1 - 5\cos^2 \alpha_j)\right]~.
\label{eq:phi}
\end{equation}

The subindices `tidal,$j$' and `rot,$j$' in eq.(\ref{eq:wtot}) denote
the tidal and rotational contributions of component $j$, and `GR'
represents the relativistic contribution to the periastron advance.
The angle $i$ is the measured inclination of the orbital plane
relative to the line of sight (see Table~\ref{tab:dimensions}),
$\alpha_j$ are the angles between the rotation axes and the normal to
the orbital plane, and $\beta_j$ are the angles between the rotation
axes and the line of sight to the observer.  The symbols $\Omega_j$
and $\Omega_K$ represent the rotational angular velocity for component
$j$, and the Keplerian orbital angular velocity; $A$ is the semimajor
axis of the orbit, and $R_j$ and $m_j$ are the radius and mass of
component $j$. The symbols $k_{2j}$ represent the apsidal motion
constant of component $j$, which we determined in the previous
section.  The auxiliary functions $f(e)$ and $g(e)$ depend only on the
eccentricity $e$, and can be written as $f(e) = (1 - e^2)^{-2}$ and $
g(e) = (8 + 12e^2 + e^4) f(e)^{2.5}/8$. The factors $\phi_j$ capture
the dependence on the orientation of the stars. When the spin axes are
perfectly aligned with the orbit, $\alpha_j = 0$, $\beta_j = i$, and
$\phi_j = 1$, so that eq.(\ref{eq:wtot}) reduces to the classical
expression presented by Sterne (1939a, 1939b).

The relativistic contribution to the apsidal motion, which is
independent of stellar structure, was given by Levi-Civita (1937) as
\begin{equation}
 \dot\omega_{\rm GR} = 2.29\times 10^{-3} {{(m_1 + m_2)}\over {A(1
 - e^2)}}~{\rm deg~cycle}^{-1}~,
\label{eq:wgr}
\end{equation}
where $m_i$ and $A$ are in solar units.

Introducing numerical values in eq.(\ref{eq:wtidal}), we obtain for
the tidal contribution $\dot\omega_{\rm tidal,1} +\dot\omega_{\rm
tidal,2} = 0.000339^{+40}_{-34}$ deg~cycle$^{-1}$, where the error
bars are given in units of the last significant digit and account for
all observational uncertainties. This is about 10\% smaller than
obtained by Albrecht et al. (2009), who adopted slightly different
absolute dimensions for the stars. The relativistic contribution is
$\dot\omega_{\rm GR} = 0.000677^{+6}_{-7}$ deg~cycle$^{-1}$.

The calculation of the rotational contribution is more involved due to
the dependence on the orientation of the system. The angles $\alpha_j$
and $\beta_j$ in eq.(\ref{eq:phi}) are not directly measured.
However, they are related through the expression
\begin{equation}
 \cos \alpha_j = \cos \beta_j \cos i + \sin \beta_j \sin i \cos \lambda_j~,
\label{eq:alpha}
\end{equation}
in which $\lambda_j$ are accessible to observation through the
Rossiter-McLaughlin effect, and represent the angles between the
projections of the spin axes and the orbital axis on the plane of the
sky, for each star. Albrecht et al.\ (2009) determined their values to
be $\lambda_1 = +72^{\circ} \pm 4^{\circ}$ and $\lambda_2 =
-84^{\circ} \pm 8^{\circ}$ for the primary and secondary of DI~Her,
respectively. The angular velocities $\Omega_j$ depend on the
projected linear rotational velocities, which were measured
spectroscopically by Albrecht et al.\ (2009) as $v_1 \sin \beta_1 =
108 \pm 4$~km~s$^{-1}$ and $v_2 \sin \beta_2 = 116 \pm 4$~km~s$^{-1}$,
and on the unknown inclination angles $\beta_j$ needed to de-project
the measured velocities in order to infer the equatorial values. We
may therefore write
\begin{equation}
 {\Omega_j\over\Omega_K} = {P\over{2\pi R_j}} {[v_j \sin \beta_j]\over{\sin \beta_j}}~,
\label{eq:Omega}
\end{equation}
in which we represent $v_j \sin \beta_j$ in square brackets to
indicate that it is a measured quantity.

We follow Albrecht et al.\ (2009) and perform Monte Carlo simulations
in order to determine $\phi_1$ and $\phi_2$, and with these the total
rate of apsidal motion in eq(\ref{eq:wtot}). The result is a
probability distribution rather than a single value. For this we
assume that the angles $\beta_j$ are distributed completely at
random\footnote{In other words, Prob($\beta_j$) d$\beta_j = \sin
\beta_j~d\beta_j$.}, or equivalently, that $\cos \beta_j$ has a
uniform distribution. Knowledge of $\beta_j$ is sufficient to
determine $\cos \alpha_j$ through eq.(\ref{eq:alpha}), and
$\Omega_j/\Omega_K$ through eq.(\ref{eq:Omega}).  As did Albrecht et
al.\ (2009), we exclude angles $\beta_j$ that lead to equatorial
rotational velocities exceeding the breakup velocity, for which we
adopt 600~km~s$^{-1}$.

\begin{figure}
\includegraphics[width=0.48\textwidth]{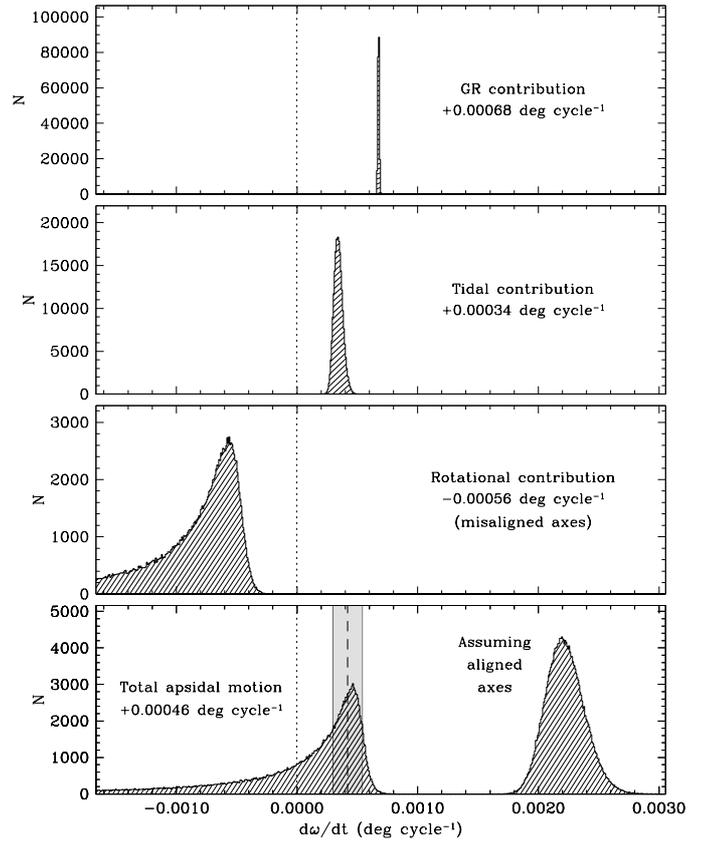}
\caption{Monte Carlo calculations for the apsidal motion of DI~Her,
based on $2 \times 10^5$ simulations. From top to bottom, the first
three panels show the separate contributions from GR, tidal
distortions (2nd and 3rd terms of eq.[\ref{eq:wtot}]), and rotational
distortions (4th and 5th terms of eq.[\ref{eq:wtot}]), based on the
measurements by Albrecht et al.\ (2009). The distributions account for
all observational errors. The bottom panel shows the predicted total
apsidal motion rate on the left (sum of all the above contributions),
and on the right the result obtained if the spin axes are assumed to
be aligned with the orbital axes, rather than strongly misaligned, as
observed. The measured rate of apsidal motion of $\dot\omega_{\rm obs}
= 0.00042 \pm 0.00012$ deg~cycle$^{-1}$ is indicated with the vertical
dashed line and shaded area, and agrees remarkably well with the
calculation for misaligned axes. The mode of the distribution for the
case of aligned axes, on the other hand, is more than 5 times larger
than the measured value.}
\label{fig:montecarlo}
\end{figure}

In the first three panels of Figure~\ref{fig:montecarlo} we show
separately the GR, tidal, and rotational contributions to the apsidal
motion rate. The distributions for the first two were also generated
with Monte Carlo simulations in which we have assumed all
observational quantities have normally distributed errors. The
distribution for the rotational term in the third panel is much wider
and asymmetrical, but also sharply peaked, and has a {\it negative}
modal value of $\dot\omega_{\rm rot} = -0.00056$ deg~cycle$^{-1}$, as
proposed by Shakura (1985) and others, indicating a retrograde motion
of the line of apsides due to the misalignment of the spin and orbital
axes. This result is numerically some 10\% larger than that obtained
by Albrecht et al.\ (2009), on account of the slightly different
absolute dimensions and theoretical $k_{2j}$ values adopted. The
bottom panel of Figure~\ref{fig:montecarlo} displays the predicted
distribution for the total apsidal motion rate (left), which has a
mode of $\dot\omega_{\rm total} = 0.000462^{+7}_{-473}$
deg~cycle$^{-1}$ (asymmetric 1$\sigma$ errors enclosing 68.3\% of the
results). The agreement with the new measured value of $\omega_{\rm
obs} = 0.00042 \pm 0.00012$ deg~cycle$^{-1}$ from Sect.~\ref{sec:wolf}
is excellent, although the theoretical distribution has a long tail
toward the left that could certainly accomodate smaller values of
$\dot\omega_{\rm obs}$. Still, it is interesting to note that the
measurement now agrees much more closely with the most probable value
from theory, the formal difference being only 10\%.  The improvement
compared to the results by Albrecht et al.\ (2009) is due mostly to
the refinement of the measured apsidal motion rate, and to some extent
also to the updated absolute dimensions and stellar models leading to
more realistic apsidal motion constants. In particular, our models
imply stars that are somewhat more centrally concentrated in mass.

If the spin axes of DI~Her are assumed to be perfectly aligned with
the orbit, instead of strongly misaligned as observed, the resulting
Monte Carlo distribution for the total apsidal motion rate is that
shown on the right-hand side of the bottom panel of
Figure~\ref{fig:montecarlo}. The mode of this distribution is at
$\dot\omega_{\rm total} = 0.00220^{+18}_{-13}$ deg~cycle$^{-1}$, which
is 5.2 times larger than the measured value. The distribution is
completely inconsistent with the observation. This result for the case
of aligned axes is nearly a factor of two larger than that presented
by Albrecht et al.\ (2009).  The reason for this significant
difference is that theirs is based on calculations by Guinan \&
Maloney (1985) in which the projected rotational velocities of both
stars were taken to be 45~km~s$^{-1}$, as measured spectroscopically
in the 1970s, whereas we used the values reported by Albrecht et al.\
(2009) themselves, obtained in 2008, which are considerably larger
(Table~\ref{tab:dimensions}).

\begin{figure}[!hb]
\includegraphics[width=0.48\textwidth]{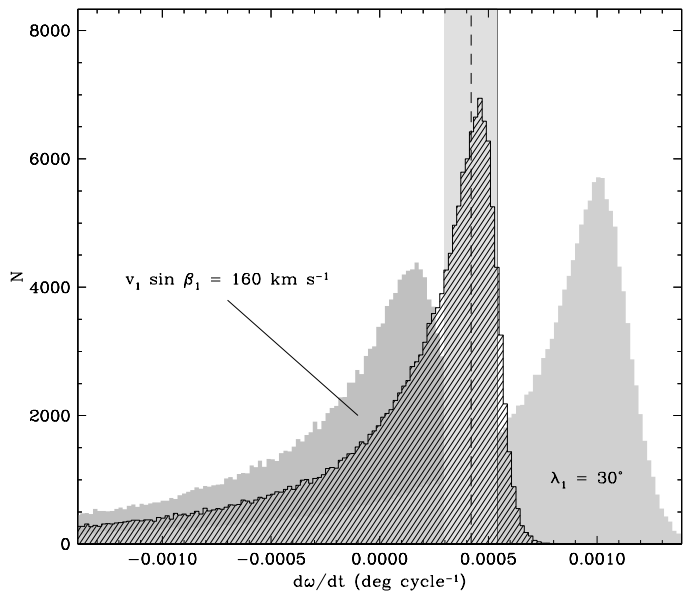}
\caption{Sensitivity of the Monte Carlo calculations for the predicted
apsidal motion rate of DI~Her to changes in some of the input
quantities measured with the Rossiter-McLaughlin effect.  The hatched
histogram is the result using the values exactly as reported by
Albrecht et al.\ (2009), and the vertical dashed line and shaded
region marks the new measured apsidal motion rate reported in this
work. The shaded histogram on the left is for $v_1 \sin \beta_1 = 160$
km~s$^{-1}$ instead of the nominal value of 108 km~s$^{-1}$, with all
other quantities left at their measured values. The shaded histogram
on the right is for $\lambda_1 = 30^{\circ}$ instead of the observed
value of 72$^{\circ}$. Similar effects are seen when varying those
parameters for the secondary.}
\label{fig:sensitivity}
\end{figure}

The sensitivity of our Monte Carlo calculations for the misaligned
case to some of the quantities measured by Albrecht et al.\ (2009),
such as $\lambda_j$ and $v_j \sin \beta_j$, is illustrated in
Figure~\ref{fig:sensitivity}. We show the changes in the resulting
distributions when these values are altered one at a time for the
primary star, with all other observational quantities held at their
nominal values.

The DI~Her observations of Albrecht et al.\ (2009) exploiting the
Rossiter-McLaughlin effect represent a breakthrough in our
understanding of this system, and the accuracy of their measurements
has played a central role in achieving the now excellent agreement
with the measured apsidal motion rate of the binary, to within 10\% (a
small fraction of the measurement error). The main factor leading to
this improvement has been the new measurement of the apsidal motion
rate presented here, followed by other refinements including the
consideration of stellar models with updated input physics, and more
accurate absolute dimensions for the binary components.  We now know
that the spin axes of the DI~Her components are strongly misaligned,
and the mystery of its slow apsidal motion rate is solved.  However,
questions remain. During the process of binary star formation the
stellar spins and the orbit presumably derive their angular momentum
from the same source, which is the parent molecular cloud. One would
then naturally expect that the axes would be closely aligned, yet in
this system they are not. Thus, interest in this object is sure to
continue as theorists try to answer these questions about DI~Her, and
observers look for other systems like it.

\begin{acknowledgements} 
The Spanish MEC (AYA2006-06375) is gratefully acknowledged for its
support during the development of this work. GT acknowledges partial
support through grant AST-0708229 from the US National Science
Foundation. The research of MW was supported by the Research Program
MSM0021620860 of the Ministry of Education of the Czech Republic.  We
would like to thank Ms.\ Hana Ku\v{c}\'akov\'a, \ova\ Observatory, and
Mr.\ Ladislav \v{S}melcer, \valmez\ Observatory, for their valuable
assistance with the photometric observations, and the anonymous
referee for helpful suggestions.  This research has made use of the
SIMBAD database, operated at CDS, Strasbourg, France, and of NASA's
Astrophysics Data System Abstract Service.
\end{acknowledgements}

\clearpage


\begin{thebibliography}{} 

\bibitem{Albrecht:09} Albrecht, S., Reffert, S., Snellen, I.\ A.\ G., \& Winn, J.\ N. 2009, \nat, 461, 373

\bibitem{Alexander:94} Alexander, D.\ R., \& Ferguson, J.\ W. 1994, \apj, 437, 879
 
\bibitem{Claret:97} Claret, A. 1997, \aap, 327, 11

\bibitem{Claret:98} Claret, A. 1998, \aap, 330, 533

\bibitem{Claret:04} Claret, A. 2004, \aap, 424, 919

\bibitem{Claret:07} Claret, A. 2007, \aap, 475, 1019

\bibitem{Company:88} Company, R., Portilla, M., \& Gim\'enez, A. 1988, \apj, 335, 962

\bibitem{Crawford:78} Crawford, D.\ L. 1978, \aj, 83,

\bibitem{Demarque:04} Demarque, P., Woo, J.-H., Kim, Y.-C., \& Yi, S.\ K. 2004, \apjs, 155, 667

\bibitem{Gimenez:83} Gim\'enez, A., \& Garc\'{\i}a-Pelayo, J.\ M. 1983, \apss, 92, 203

\bibitem{Guinan:85} Guinan, E.\ F., Maloney, F. P. 1985, \aj, 90, 1519

\bibitem{Hilditch:75} Hilditch, R.\ W., \& Hill, G. 1975, Mem.\ R.\ astr.\ Soc., 79, 101

\bibitem{Hut:81} Hut, P. 1981, \aap, 99, 126

\bibitem{Iglesias:96} Iglesias, C.\ A., \& Rogers, F. J. 1996, \apj, 464, 943

\bibitem{khodykin:07} Khodykin, S.\ A. 2007, IBVS, No.\ 5788

\bibitem{Kozyreva:09} Kozyreva, V.\ S., \& Bagaev, L.\ A. 2009, Astronomy Letters, 35, 483

\bibitem{Kwee:56} Kwee, K.\ K., \& van Woerden, H. 1956, Bull.\ Astron.\ Inst.\ Netherlands, 12, 327

\bibitem{Levi-Civita:37} Levi-Civita, T. 1937, Amer. J. Math., 59, 225

\bibitem{Motl:07} Motl, D. 2007, {\sc C-Munipack}, http://c-munipack.sourceforge.net

\bibitem{Paunzen:05} Paunzen, E., Schnell, A., \& Maitzen, H.\ M. 2005, \aap, 444, 941

\bibitem{Popper:80} Popper D.\ M.\ 1980, \araa, 18, 115

\bibitem{Popper:82} Popper D.\ M. 1982, \apj, 254, 203

\bibitem{Runkle:03} Runkle, R.\ C. 2003, Ph.D.\ Thesis, University of North Carolina, unpublished

\bibitem{Shakura:85} Shakura, N.\ I. 1985, Sov. Astr. Let., 11(4), 225

\bibitem{Sterne:39a} Sterne, T.\ E. 1939a, \mnras, 99, 451

\bibitem{Sterne:39b} Sterne, T.\ E. 1939b, \mnras, 99, 662

\bibitem{Torres:09a} Torres, G., Andersen, J., \& Gim\'enez, A. 2009a, \aapr, 18, 67

\bibitem{Torres:09b} Torres, G., Lacy, C.\ H.\ S., \& Claret, A. 2009b, \aj, 138, 1622

\bibitem{Wolf:09} Wolf, M., Claret, A., Kotkova, H., Kocian, R., Brat, L., Svoboda, P., \& Smelcer, L. 2009, \aap, in press

\bibitem{Yi:01} Yi, S.\ K., Demarque, P., Kim, Y.-C., Lee, Y.-W., Ree, C.\ H., Lejeune, T., \& Barnes, S. 2001, \apjs, 136, 417

\bibitem{Zahn:77} Zahn, J.-P. 1977, \aap, 57, 383

\bibitem{Zahn:89} Zahn, J.-P. 1989, \aap, 220, 112

\end{thebibliography}
\end{document}